\documentclass[a4paper,11pt]{article}
\pdfoutput=1 \usepackage{jheppub_2} 
\usepackage[T1]{fontenc} 
\usepackage[utf8]{inputenc}
\usepackage{color,subfigure}
\usepackage{graphicx, multirow,soul,url,amsmath,amsfonts,amssymb,mathrsfs,amsfonts}
\usepackage[all]{hypcap}
\usepackage{url}
\usepackage{bbm}
\usepackage{cancel}
\usepackage{subfigure}
\usepackage{slashed}
\usepackage[dvipsnames]{xcolor}
\usepackage{multicol, blindtext}
\usepackage[section]{placeins}
\usepackage[version=3]{mhchem}
\usepackage{caption}
\usepackage{lipsum}
\usepackage{hyperref}
\usepackage{float}
\usepackage{mattens}
\usepackage{mathtools}
\usepackage{footnote}
\usepackage{amsmath,amssymb,amsthm,amsxtra,overpic,bbm,bm,epsfig}
\usepackage{color,subfigure}
\usepackage[splitrule]{footmisc}
\usepackage{bbold}
\usepackage{amsmath}
\usepackage[normalem]{ulem}

\newcommand{\beq}{\begin{equation}}
\newcommand{\eeq}{\end{equation}}

\newcommand{\SU}{\,{\rm SU}}

\newcommand{\U}{\,{\rm U}}

\newcommand{\matrixel}[3]{\left< #1 \vphantom{#2#3} \right|
 #2 \left| #3 \vphantom{#1#2} \right>} 
\let\baraccent=\= 

\notoc 

\title{\boldmath {{Axion Dark Matter, Proton Decay and Unification}}}
\author[a]{Pavel Fileviez P\'erez,}
\author[b]{Clara Murgui}
\author[a]{and Alexis D. Plascencia}
\affiliation[a]{Physics Department and Center for Education and Research in Cosmology and Astrophysics (CERCA), 
Case Western Reserve University, Cleveland, OH 44106, USA}
\affiliation[b]{Departamento de F\'isica Te\'orica, IFIC, Universitat de Valencia-CSIC, 
E-46071, Valencia, Spain}

\emailAdd{pxf112@case.edu}
\emailAdd{clara.murgui@ific.uv.es}
\emailAdd{alexis.plascencia@case.edu}

\abstract{
We discuss the possibility to predict the QCD axion mass in the context of grand unified theories. We investigate the implementation of the DFSZ mechanism in the context of renormalizable SU(5) theories. 
In the simplest theory, the axion mass can be predicted with good precision in the range $m_a = (2-16)$ neV, and there is a strong correlation between the predictions for the axion mass and proton decay rates. In this context, we predict an upper bound for the proton decay channels with antineutrinos, $\tau(p\to K^+ \bar{\nu}) \lesssim   4 \times 10^{37} \text{ yr}$ and $\tau(p \to \pi^+ \bar{\nu}) \lesssim 2 \times 10^{36}\text{ yr}$. This theory can be considered as the minimal realistic grand unified theory with the DFSZ mechanism and it can be fully tested by proton decay and axion experiments.}

\begin{document} 

\maketitle
\flushbottom

\newpage 

\section{Introduction}
The QCD axion~\cite{Peccei:1977hh,Wilczek:1977pj,Weinberg:1977ma} is one of the most motivated dark matter candidates present in theories for physics beyond the Standard Model (SM). 
The existence of the axion field is motivated by two of the shortcomings of the SM; namely, the strong CP problem\footnote{Recently, we had a wonderful discussion about the strong CP problem with Goran Senjanovi\'c, where he showed us that the strong CP problem might not be a severe problem. These results have not been published and here we just refer to our discussions. However, the PQ mechanism still remains an appealing dynamical explanation for the smallness of $\bar{\theta}$.} and the existence of dark matter in the Universe~\cite{Preskill:1982cy,Abbott:1982af,Dine:1982ah}.
Unfortunately, since its mass and interactions are determined by the unknown Peccei-Quinn (PQ) symmetry breaking scale it is difficult to make predictions for the experiments. 
For reviews about the axion we refer the reader to Refs.~\cite{Raffelt:1990yz,Dine:2000cj,Sikivie:2006ni,Kim:2008hd,Jaeckel:2010ni,Marsh:2015xka,Graham:2015ouw,Irastorza:2018dyq}.

Dine, Fischler, Srednicki and Zhitnitsky (DFSZ)~\cite{Zhitnitsky:1980tq, Dine:1981rt} proposed a simple mechanism where the origin of the axion mass and its couplings can be understood. In this scenario, a second Higgs doublet and a new scalar singlet are added to the SM. The axion field lives mostly in the electroweak (EW) singlet and it is coupled indirectly to the fermions in the SM through mixing terms in the potential. Therefore, by performing a chiral rotation of the quark fields the coupling between the axion and the gluons, $aG\tilde{G}$, can be generated. A second simple model for the QCD axion was proposed by Kim, Shifman, Vainshtein, and Zakharov (KSVZ) in Refs.~\cite{Kim:1979if,Shifman:1979if} where after integrating extra colored matter one generates the $aG\tilde{G}$ term. Unfortunately, these axion models do not provide any information about the PQ scale. In order to predict the axion mass one needs to connect the PQ scale with the scale of new physics predicted in a given theory.
Recently, we have discussed in Ref.~\cite{FileviezPerez:2019fku} the simplest renormalizable grand unified theory where the KSVZ mechanism for the axion can be implemented and predicted the axion mass using the fact that the PQ scale is defined by the unification scale. For other studies of axions in GUTs, see Refs.~\cite{Co:2016vsi,Boucenna:2017fna,DiLuzio:2018gqe,Ernst:2018bib}.

In this article, we study the implementation of the DFSZ in the context of a renormalizable $\SU(5)$ grand unified theory. Following the original idea of Wise, Georgi and Glashow~\cite{Wise:1981ry} we promote the $\mathbf{24_H}$ to a complex field by imposing a global $\U(1)_{\rm PQ}$ symmetry. The fact that its vacuum expectation value breaks simultaneously $\SU(5)$ and $\U(1)_{\rm PQ}$ establishes a connection between the PQ and the GUT scale. 
However, the scenario discussed in Ref.~\cite{Wise:1981ry} is ruled out by the experiment and here we will study the predictions for the axion mass and couplings in a realistic renormalizable grand unified theory.
We find that the axion mass is predicted to be in the window $m_a = (2-16)$ neV, range that the ABRACADABRA~\cite{Kahn:2016aff}  experiment in combination with the CASPEr-Electric~\cite{Budker:2013hfa} experiment will be able to fully probe. This theory can be considered as the simplest realistic grand unified theory where the DFSZ mechanism is implemented and a strong correlation between the axion mass and the proton decay rates is predicted.
In this context one can predict upper bounds on the proton decay lifetimes for the channels with antineutrinos, i.e. $\tau(p\to K^+ \bar{\nu}) \lesssim   4 \times 10^{37} \text{ yr}$  and  
$\tau(p \to \pi^+ \bar{\nu}) \lesssim 2 \times 10^{36}\text{ yr}$. Therefore, the theory we discuss in this article can be fully tested in the near future at proton decay and axion experiments.

This article is organized as follows: in section~\ref{sec:theory} we describe a simple GUT that allows for a global PQ symmetry and is realistic, in the sense that reproduces the values of the gauge couplings at the electroweak scale and can explain the masses for the charged fermions. In section~\ref{sec:DFSZ} we discuss how to implement the DFSZ mechanism in the context of $\SU(5)$ GUTs and establish a direct connection between the Peccei-Quinn scale and the GUT scale. We also demonstrate that this theory is able to provide predictions for the axion mass and the proton lifetime in the channel involving anti-neutrinos. In section~\ref{sec:axionpheno}  we discuss the testability of the theory by studying the axion-photon coupling and the axion coupling to the neutron electric dipole moment in the predicted mass window.

\section{Theoretical Framework}
\label{sec:theory}
In order to predict the axion mass via the DFSZ mechanism, we work with the renormalizable $\SU(5)$ grand unified theory where the matter fields of the SM are unified 
in $\mathbf{\bar{5}}$ and $\mathbf{10}$ representations. The Higgs sector is composed of the minimal representations required for the spontaneous symmetry breaking of the theory, $\mathbf{5_H}$ and $\mathbf{24_H}$, 
\begin{eqnarray*}
 \displaystyle {\mathbf{5_H}}  &\sim& \underbrace{(1,2,1/2)}_{H_1} \oplus \underbrace{(3,1,-1/3)}_{T}, \\
 \displaystyle  {\mathbf{24_H}} &\sim& \underbrace{(8,1,0)}_{\Sigma_8} \oplus \underbrace{(1,3,0)}_{\Sigma_3} \oplus \underbrace{(3,2,-5/6)}_{\Sigma_{(3,2)}} \oplus \underbrace{(\bar{3},2,5/6)}_{\Sigma_{(\bar{3},2)}} \oplus \underbrace{(1,1,0)}_{\Sigma_0},
\end{eqnarray*}
and a $\mathbf{45_H}$ needed to correct the mass relation between down-type quarks and charged leptons
\begin{equation*}
 \displaystyle {{\mathbf{45_H}}} \sim\underbrace{(1,2,1/2)}_{H_2} \oplus \underbrace{(8,2,1/2)}_{\Phi_1}\oplus \underbrace{(\bar{6},1,-1/3)}_{\Phi_2}\oplus \underbrace{(3,3,-1/3)}_{\Phi_3}
 \oplus \underbrace{(\bar{3},2,-7/6)}_{\Phi_4}\oplus \underbrace{(3,1,-1/3)}_{\Phi_5} \oplus \underbrace{(\bar{3},1,4/3)}_{\Phi_6}.
\end{equation*}
In order to implement the PQ mechanism we impose a $\U(1)_\text{PQ}$; in this context the $\mathbf{24_H}$ becomes complex and allows for a CP-odd field that will become the axion after the global symmetry is spontaneously broken. Then, the axion lives mostly in ${\bf{24_H}}$, i.e. $${\bf{24_H}} \supset \frac{1}{\sqrt{2}} |\Sigma_0| e^{i a(x)/v_\Sigma},$$ with $v_\Sigma$ being the vacuum expectation value of $\Sigma_0$. We note that a mixing term between all the CP-odd Higgses cannot be generated because $\mathbf{45_H}$ and $\mathbf{5_H}$ must be equally charged under PQ in order to correct the charged fermion masses. Following the approach by Wise, Georgi and Glashow~\cite{Wise:1981ry}, the aforementioned problem is solved by adding an extra Higgs in the fundamental representation
\begin{equation*}
 \displaystyle \mathbf{5_H^{'}} \sim \underbrace{(1,2,1/2)}_{H_3} \oplus \underbrace{(3,1,-1/3)}_{T^{'}}, 
 \end{equation*}
whose mixing term with the CP-odd phase in $\mathbf{24_H}$,
\beq
\label{eq:DFSZterms}
V \supset   5_H'^\dagger \, 24_H^2  ({\lambda}_1 \, 5_H + {\lambda}_2 \, 45_H) + {\lambda}_3 \, 5_H'^{\dagger} \, 5_H \,  {\rm Tr}[24_H^2]  +  {\rm h.c.},
\eeq
allows for the implementation of the DFSZ mechanism once the symmetry is spontaneously broken. In this context, the $\SU(5)\otimes \U(1)_\text{PQ}$ theory has the following Yukawa interactions:
\begin{equation}
\label{eq:Yukawas}
{\cal L}_Y = Y_u \ 10 \ 10 \ 5_H^{'} \ + \ 10 \ \bar{5} \ \left( Y_1 \ 5_H^* \ + \ Y_2 \ 45_H^* \right) \ + \  {\rm{h.c.}}.
\end{equation}
Since the above terms must respect the global PQ symmetry, it follows that the $\U(1)_{\rm PQ}$ charges are given by
\begin{center}
\begin{tabular}{r | c c c c c c }
Field 		& $\bar{5}$		&  $10$ 		&  $5_H^{'} $  	& $ 5_H $			& $45_H$			& 	$24_H$ \\
\hline
PQ charge 	&  $\alpha$		&  $\beta$		& $-2\beta$	& $\alpha + \beta$	& $\alpha + \beta$	& $-(\alpha+3\beta)/2$
\end{tabular}
\end{center}
where the PQ charge of $\mathbf{24_H}$ is determined by the potential in Eq.~\eqref{eq:DFSZterms}, which defines the mixing between the axion and the Higgs doublets. 

In this theory the mass matrices for the charged fermions are given by
\begin{eqnarray}
M_u &=& \sqrt{2} \left( Y_u + Y_u^T \right) v_{5'}, \\
M_e &=& \frac{1}{2} \left( Y_1 v_5^* - 6 Y_2 v_{45}^* \right), \\
M_d &=& \frac{1}{2} \left( Y_1^T v_5^* + 2 Y_2^T v_{45}^* \right).
\end{eqnarray}
Notice that $M_u=M_u^T$ has strong implications for the proton decay channels with antineutrinos~\cite{FileviezPerez:2004hn}. In Appendix~\ref{sec:AppNu} we discuss how to extend this model to explain neutrino masses.
%

\section{The DFSZ mechanism in SU(5)}
\label{sec:DFSZ}
%
The terms in the scalar potential relevant for the DFSZ mechanism can be written in terms of the elements of the $\SU(5)$ representations in the following way
\begin{equation}
V  \supset  (m_{12}^2  + \lambda H_1^\dagger H_2)  H_1^\dagger H_2 +H_3^\dagger \Sigma_0^2 ( \lambda_\text{a1} H_1 + \lambda_\text{a2} H_2)+ \text{h.c.},
\end{equation}
where, following the notation previously introduced, $H_1 \subset \mathbf{5_H}$, $H_2 \subset \mathbf{45_H}$, $H_3 \subset \mathbf{5'_H}$ and $\Sigma_0 \subset \mathbf{24_H}$. After spontaneous symmetry breaking, all neutral fields acquire a vacuum expectation value  and one can write their CP-odd component as a function of the two Goldstone bosons arising due to the spontaneous breaking of two global symmetries of the potential:
\begin{eqnarray}
H_1^0 &\supset&   \frac{v_1}{\sqrt{2}}  e^{i \, a_1 / v_1} = \frac{v_1}{\sqrt{2}} e^{i (q\, \hat{a}_Z + \text{PQ}_1 \, \hat{a} )}, \label{eq:Goldstone1} \\
H_2^0 &\supset&   \frac{v_2}{\sqrt{2}}  e^{i \, a_2 / v_2} = \frac{v_2}{\sqrt{2}} e^{i (q \, \hat{a}_Z + \text{PQ}_2 \, \hat{a} )},\\
H_3^0 &\supset&   \frac{v_3}{\sqrt{2}}  e^{i \, a_3 / v_3} = \frac{v_3}{\sqrt{2}} e^{i (q \, \hat{a}_Z + \text{PQ}_3 \, \hat{a} )},\\
\Sigma_0	 &  \supset&   \frac{v_\Sigma}{\sqrt{2}}  e^{i \, a_\Sigma / v_\Sigma} = \frac{v_\Sigma}{\sqrt{2}} e^{i \,  \text{PQ}_\Sigma \, \hat{a} }. \label{eq:Goldstone2}
\end{eqnarray}
Here, $\hat{a}$ and $\hat{a_Z}$ are the phases of the axion and the Goldstone boson that will be eaten by the $Z$, respectively. The factor $q$ is the contribution related to the electroweak quantum numbers and $\text{PQ}_i$ parametrizes the presence of the axion in each of the scalar representations.
The terms of the scalar potential fix the following conditions for the ${\rm PQ}_i$ charges:
\begin{equation}
\text{PQ}_1 = \text{PQ}_2 , \quad \text{ and } \quad \text{PQ}_{1,2} + 2 \text{PQ}_\Sigma - \text{PQ}_3 = 0.
\end{equation}
Notice that the axion in reality is a pseudo-Goldstone boson since the Peccei-Quinn symmetry, although being a good symmetry classically, it is broken at the quantum level.
Linearizing the kinetic terms,
\begin{eqnarray}
\frac{1}{2}\partial_\mu a_1 \partial^\mu a_1 &=& \frac{1}{2} v_1^2 \left( q^2 \partial_\mu \hat{a}_Z \partial^\mu \hat{a}_Z + 2 q \, \text{PQ}_1 \,  \partial_\mu \hat{a}_Z \partial^\mu \hat{a} + \text{PQ}_1^2 \, \partial_\mu \hat{a} \partial^\mu \hat{a} \right),\\
\frac{1}{2}\partial_\mu a_2 \partial^\mu a_2 &=& \frac{1}{2} v_2^2 \left( q^2 \partial_\mu \hat{a}_Z \partial^\mu \hat{a}_Z + 2 q \, \text{PQ}_2 \, \partial_\mu \hat{a}_Z \partial^\mu \hat{a} + \text{PQ}_2^2 \, \partial_\mu \hat{a} \partial^\mu \hat{a} \right),\\
\frac{1}{2}\partial_\mu a_3 \partial^\mu a_3 &=& \frac{1}{2} v_3^2 \left( q^2 \partial_\mu \hat{a}_Z \partial^\mu \hat{a}_Z + 2 q \, \text{PQ}_3 \,  \partial_\mu \hat{a}_Z \partial^\mu \hat{a} + \text{PQ}_3^2  \, \partial_\mu \hat{a} \partial^\mu \hat{a} \right),\\
\frac{1}{2}\partial_\mu a_\Sigma \partial^\mu a_\Sigma &=& \frac{1}{2} v_\Sigma^2 \, \text{PQ}_\Sigma^2 \, \partial_\mu \hat{a} \partial^\mu \hat{a} .
\end{eqnarray}
Orthogonality of the Goldstone bosons requires the following condition
\begin{equation}
v_1^2\, \text{PQ}_1 +v_2^2 \, \text{PQ}_2 +v_3^2 \, \text{PQ}_3 =  0,
\end{equation}
whereas the normalization of the kinetic terms of the axion demands that
\begin{equation}
v_1^2 \, \text{PQ}_1^2 + v_2^2 \,  \text{PQ}_2^2 + v_3^2  \, \text{PQ}_3^2 + v_\Sigma^2 \, \text{PQ}_\Sigma^2 = n_a^2,
\end{equation}
where $n_a$ is the normalization of the CP-odd phase $\hat{a} = a / n_a$. The presence of the axion in each of the scalar representations is given by
\begin{equation}
\label{eq:axionrep}
H_1^0  \supset  \frac{v_1}{\sqrt{2}} e^{i \frac{a}{n}}, \quad H_2^0 \supset \frac{v_2}{\sqrt{2}} e^{i  \frac{a}{n}}, \quad 
H_3^0  \supset  \frac{v_3}{\sqrt{2}} e^{-i\left(\frac{v_1^2 + v_2^2}{v_3^2}\right)\frac{a}{n}}, \quad  \Sigma_0  \supset \frac{v_\Sigma}{\sqrt{2}} e^{-i \left(\frac{v^2}{2 v_3^2} \right) \frac{a}{n}},
\end{equation}
where $v \equiv \sqrt{v_1^2 + v_2^2 + v_3^2} = 246 \text{ GeV}$ and 
\begin{equation}
n \equiv \frac{n_a}{\text{PQ}_1} =  \frac{\sqrt{v_1^2 + v_2^2} }{v_3} \, v \, \displaystyle \sqrt{1+ \frac{v_\Sigma^2 \, v^2}{4 v_3^2 (v_1^2 + v_2^2)}} \simeq \frac{v^2 \, v_\Sigma}{2 \, v_3^2},
\end{equation}
for the last relation we have taken the limit $v_\Sigma \gg v_1, v_2, v_3$, which is justified by the fact that $M_{\rm GUT}$ is about 13 orders of magnitude higher than the electroweak scale. Comparing Eqs.~\eqref{eq:Goldstone1}-\eqref{eq:Goldstone2} with Eq.~\eqref{eq:axionrep} we conclude that the axion lives predominantly in the $\Sigma_0$ field as expected.

In the broken phase, the Yukawa Lagrangian can be rewritten as
\begin{equation}
{\cal L}_Y \supset M_u \,  \bar{u}_R  u_L \,  e^{-i \left( \frac{v_1^2+v_2^2}{v_3^2} \right) \frac{a}{{n}}}+ M_d \, \bar{d}_R  d_L \, e^{-i \frac{a}{{n}}} + M_e \, \bar{e}_R  e_L \, e^{-i \frac{a}{{n}}},	
\end{equation}
where the axion can be rotated away from the Yukawa Lagrangian by performing the following chiral rotations
\begin{equation}
u_{L/R} \to e^{\pm i \left( \frac{v_1^2+v_2^2}{2v_3^2} \right)\frac{a}{{n}}} u_{L/R}, \quad d_{L/R} \to e^{\pm i \frac{a}{2 {n}}}d_{L/R},  \quad \text{ and }\quad e_{L/R} \to e^{\pm i \frac{a}{2 {n}}} e_{L/R},
\end{equation}
the transformation of the quarks will generate the following $a \tilde{G}G$ term
\beq
\mathcal{L} \supset \frac{\alpha_3} {8 \pi} \frac{a}{f_a} G_{\mu\nu} \tilde{G}^{\mu\nu} = \frac{\alpha_3} {8 \pi} \frac{a}{v_\Sigma} N G_{\mu\nu} \tilde{G}^{\mu\nu} = \frac{\alpha_3} {8 \pi} \frac{a}{{n}} \frac{3v^2}{v_3^2} G_{\mu\nu} \tilde{G}^{\mu\nu}.
\eeq
Hence, the Peccei-Quinn scale $f_a$ is identified as
\begin{equation}
f_a \equiv \frac{v_\Sigma}{N} = \frac{ v_3^2 }{ 3 v^2 } {n} \simeq \frac{v_\Sigma}{6}.
\end{equation}
Then, the connection between the PQ and GUT scales is given by
\beq
f_a =\frac{M_{\rm GUT}}{\sqrt{\alpha_{\rm GUT}}} \,  \frac{1}{2 \sqrt{30\pi}},
\eeq
using $M_V\!=\!M_{\rm GUT}\!=\! \sqrt{(10  \pi/3 ) \,  \alpha_\text{GUT} } \, v_\Sigma$, where $M_V$ refers to the mass of the heavy gauge bosons mediating proton decay. 
For the relation between $m_a$ and $f_a$ we use the recent results from Ref.~\cite{Gorghetto:2018ocs}
\begin{equation}
m_a=5.691(51) \, \times 10^{-6} \ \text{eV} \left(\frac{10^{12}\text{ GeV}}{f_a}\right). 
\end{equation} 
Therefore, if we predict the GUT scale and $\alpha_{\rm GUT}$ we can determine the allowed values for the axion mass in this grand unified theory. 
%
\subsection{Unification Constraints}
\label{sec:Unification}
The fact that $\langle \mathbf{24_H} \rangle$ breaks both SU(5) and $\U(1)_\text{PQ}$ symmetries establishes a connection between the PQ and the GUT scales. Therefore, once $M_\text{GUT}$ and $\alpha_{\rm GUT}$ are known the axion mass is predicted. In this theory, $M_\text{GUT}$ is determined from the experimental input on the values of the gauge couplings at the low scale. The following RG equations fix the EW values for the gauge couplings as a function of the GUT parameters:
\begin{equation}
 \alpha_i^{-1} (M_Z) =  \alpha_\text{GUT}^{-1}  + \frac{1}{2 \pi} b_i^{\rm SM} \ {\rm{ln}} \frac{M_\text{GUT}}{M_Z} + \frac{1}{2 \pi} \sum_I b_{iI} \ \Theta (M_\text{GUT} - M_I) {\rm{ln}} \frac{M_\text{GUT}}{M_I},
 \label{RGE}
\end{equation}
where the subindex $i=1,2 \text{ and }3$ refers to the three different SM forces $\U(1)_Y$, $\SU(2)_L$ and $\SU(3)_C$, respectively, $b_1^{\rm SM}=41/10$, $b_2^{\rm SM}=-19/6$ and $b_3^{\rm SM}=-7$, and $M_I$ is the mass of any intermediate field between the EW and the GUT scales. Besides the field content of the Georgi and Glashow SU(5), this theory contains new Higgses, $\mathbf{5_H'}$ and $\mathbf{45_H}$, which may contribute to the evolution of the couplings according to Eq.~\eqref{RGE}. In Table~\ref{tab:beta} we show their contributions to the beta functions.
\begin{table}[h]
\label{tab:beta}
\centering
\begin{tabular}{c | c c c c c c c c c }

\text{Fields} 
		 	& $H_2$ 	& $H_3$	& $T'$	&  $\Phi_1$	& $\Phi_2$ 	& $\Phi_3$	& $\Phi_4$	& $\Phi_5$	& $\Phi_6$	\\
		\hline
		\hline		
$b_1$  		& 1/10 	& 1/10		&1/15	&	4/5		&	2/15		&	1/5		&	49/30	&	1/15		&	16/15	\\
		\hline
$b_2$	& 1/6		& 1/6		& 0			&	4/3		&	0		&	2		&	1/2		&	0		&	0		\\
		\hline
$b_3$	& 0		& 0			& 1/6		&	2		&	5/6		&	1/2		&	1/3		&	1/6		&	1/6		\\
		\hline
		\hline
$B_{12} / r_i$ 	& -1/15  & -1/15 & 1/15 & -8/15 & 2/15 & -9/5  & 17/15 & 1/15  & 16/15  \\
\hline
$B_{23} / r_i$ 	& 1/6  & 1/6  & -1/6  & -2/3  & -5/6  & 3/2  & 1/6  & -1/6 & -1/6  \\
\hline
\end{tabular}
\caption{\label{tab:beta} Contributions of the new scalar sector to the running of the gauge couplings.}
\end{table}

In order to obtain the following parameters at the $M_Z$ scale: $\sin^2 \theta_W (M_Z)=0.23122$, $\alpha (M_Z)=1/127.955$, and $\alpha_s (M_Z)=0.1181$~\cite{Tanabashi:2018oca}, the following conditions must be satisfied~\cite{Giveon:1991zm}
\begin{equation}
\frac{B_{23}}{B_{12}} = 0.717, \quad \text{ and }\quad {\rm{ln}}  \frac{M_{\rm GUT}}{M_Z} =\frac{184.95}{B_{12}},
\label{unificationCONS2}
\end{equation}
where unification at the one-loop level has been assumed. The $B_{ij} \equiv B_i - B_j$, where $B_i$ is defined as
\begin{equation}
B_i \equiv b_i^\text{SM} + \sum_I b_i^I \, r^I, \, \quad \text{and }\quad  \, r^I \equiv \frac{\text{ln} (M_\text{GUT}/M_I)}{\text{ln}(M_\text{GUT}/M_Z)},
\end{equation}
are only sensitive to the relative splitting between the $\SU(5)$ representations. 

In Tab.~\ref{tab:beta} we also show their $B_{ij}$ contributions. Among the new scalar sector, the $\Phi_3$ and $H_2$ from the $\mathbf{45_H}$, together with $H_3$ in the $\mathbf{5_H'}$ help towards unification since the three of them contribute to enhance the ratio $B_{23}/B_{12}$. This helps to satisfy the first condition from Eq.~\eqref{unificationCONS2} since, as it is well known, in the Georgi and Glashow model this ratio is below the required value. The colored octet $\Phi_1$ in the $\mathbf{45_H}$ indirectly helps to unify since it allows for a larger  $M_\text{GUT}$ range according to the second condition from Eq.~\eqref{unificationCONS2}. We will assume the rest of the scalar fields to be at the GUT scale since they do not help to achieve unification.

\begin{figure}[h]
\centering
\includegraphics[height=7cm]{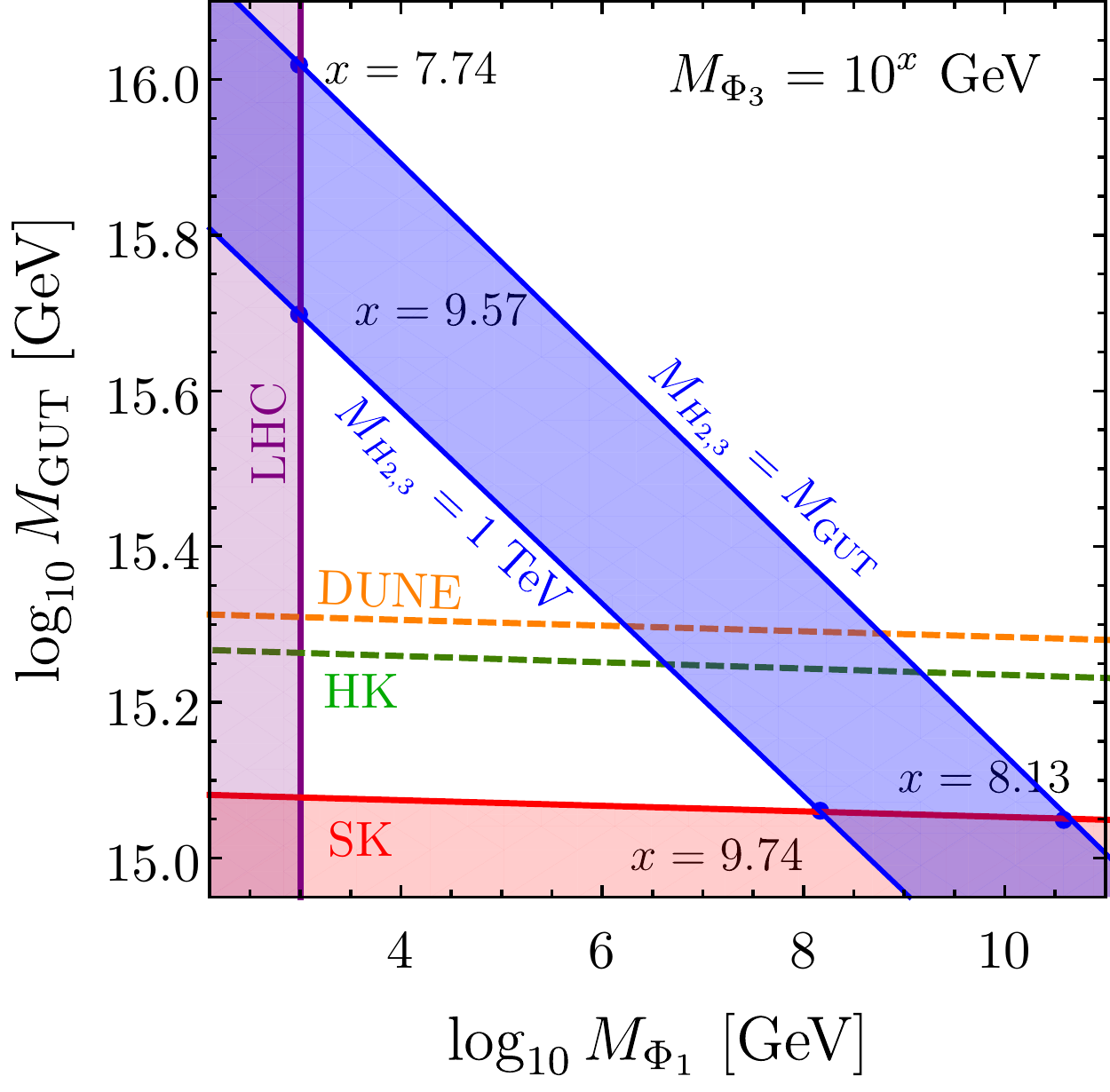}
\includegraphics[width=0.495\linewidth]{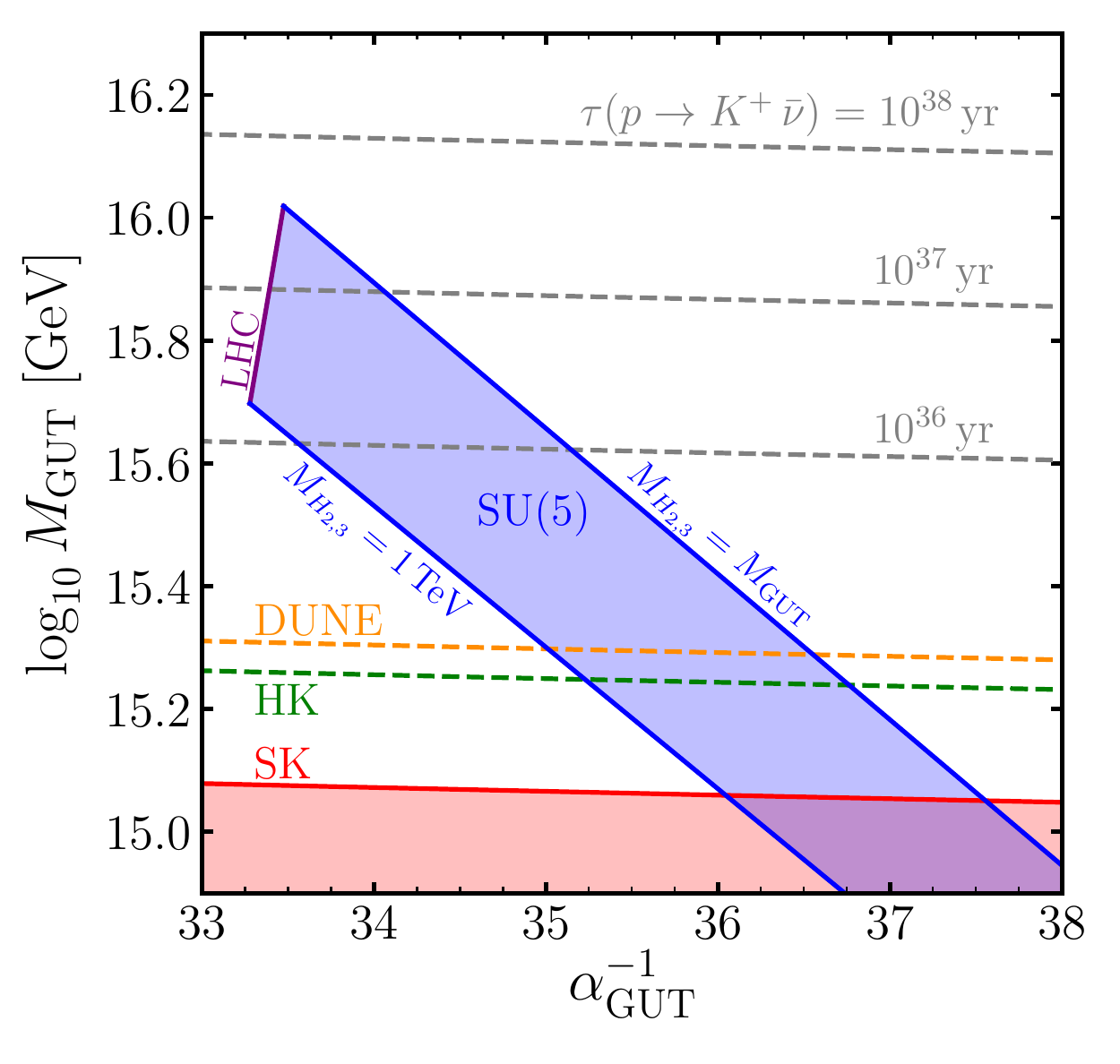}
\caption{ Unification constraints in the parameter space. The region shaded in blue satisfies the unification constraints. The red area corresponds to the parameter space ruled out by the Super-Kamiokande (SK) collaboration $\tau(p \to K^+ \bar{\nu}) > 5.9 \times 10^{33}$ years~\cite{Abe:2014mwa}, the green and orange dashed lines show the projected bounds from the Hyper-Kamiokande (HK) collaboration $\tau(p\to K^+ \bar{\nu}) > 3.2 \times 10^{34}$ years~\cite{Abe:2018uyc} and the DUNE collaboration $\tau(p\to K^+ \bar{\nu}) > 5 \times 10^{34}$ years~\cite{Acciarri:2015uup}, respectively. The parameter space excluded by collider bounds on $M_{\Phi_1}>1$ TeV~\cite{Miralles:2019uzg} is colored in purple. {\textit{Left Panel:}} In blue, prediction of the GUT scale as a function of the $\Phi_1 \sim (8,2,1/2)$ mass. {\textit{Right Panel:}} In blue, prediction of the GUT scale as a function of $\alpha_\text{GUT}$. In gray dashed lines we show the sensitivity of different hypothetical decay widths for the $p \to K^+ \bar{\nu}$ channels. In both panels the width of the blue band scans over the possible mass range for the Higgs doublets $H_2$ and $H_3$. }
\label{fig:unification}
\end{figure}

Unification constraints determine $M_\text{GUT}$ as a function of $M_{\Phi_1}$ and the doublet masses $M_{H_2}$ and $M_{H_3}$, as shown by the blue region in the left panel in Fig.~\ref{fig:unification}. This figure also shows that the lighter $\Phi_1$, the larger $M_\text{GUT}$ can be. However, due to experimental bounds derived from collider physics, $\Phi_1$ cannot be arbitrarily light. According to the recent study in Ref.~\cite{Miralles:2019uzg}, its mass has to be above 1 TeV, which establishes an upper bound for the GUT scale as the figure reflects. The region shaded in purple shows the parameter space ruled out by the collider bounds on $\Phi_1$. The mass of the $\Phi_3\sim (3,3,-1/3)$ is also obtained from the unification constraints and it is implicitly given in the figure: for $M_{H_{2,3}} = M_\text{GUT}$, it ranges from $M_{\Phi_3} \subset [5.5,\, 13.5] \times 10^{7}$ GeV, whereas for $M_{H_{2,3}} = 1$ TeV, $M_{\Phi_3} \subset [3.7,\, 5.5] \times 10^{9}$ GeV, as shown explicitly by the blue dots.

On the right panel of Fig.~\ref{fig:unification} we show the relation between $M_\text{GUT}$ and $\alpha_\text{GUT}$. The region shaded in blue satisfies the unification constraints in Eq.~\eqref{unificationCONS2}, as we vary $M_{H_2}$ and $M_{H_3}$ from 1 TeV to the GUT scale. The LHC bound on $\Phi_1$ is shown by a purple line. Experimental constraints for proton decay define a lower bound on the GUT scale. In both of the panels in Fig.~
\ref{fig:unification}, we show in red the excluded region by the bound on $p \to K^+ \bar{\nu}$ from the Super-Kamiokande (SK) collaboration, $\tau(p \to K^+ \bar{\nu}) > 5.9 \times 10^{33}$ years~\cite{Abe:2014mwa}, whereas the projected bound on the same decay from the Hyper-Kamiokande (HK) collaboration, $\tau(p\to K^+ \bar{\nu}) > 3.2 \times 10^{34}$ years~\cite{Abe:2018uyc} and the DUNE collaboration, $\tau(p\to K^+ \bar{\nu}) > 5 \times 10^{34}$ years~\cite{Acciarri:2015uup}, are shown with a green and orange dashed lines, respectively. These constraints will be addressed in detail in the next section.  

To close this section, we emphasize that with the $\mathbf{45_H}$ alone unification can be achieved, as shown in Fig.~\ref{fig:unification}, and the splitting in the $\mathbf{5_H'}$ helps to increase the parameter space where unification occurs. We find that the allowed window for the GUT scale is given by
\begin{equation}
M_\text{GUT} = (1.12 \, - \, 10.45 ) \times 10^{15} \text{ GeV},
\end{equation}
where the upper bound is obtained from the collider bounds on $M_{\Phi_1}$ whereas the lower bound is given by experimental constraints on proton decay. We note that there are two upper (and lower) bounds for the GUT scale, depending on the masses of the doublets $H_2$ and $H_3$. In order to define the GUT scale window, we have taken the conservative approach to consider the larger range possible in the context of this theory.
We also note that the consistency with proton decay bounds in this case is ensured by the PQ symmetry, because it forbids the interaction of the scalar leptoquark $\Phi_3$  with the up-quarks. Otherwise, $\Phi_3$ would mediate proton decay interactions through an effective operator suppressed by two powers of its mass. Hence, in this scenario the PQ symmetry allows $\Phi_3$ to be light in agreement with both unification constraints and proton decay.

It is well-known that in any grand unified theory one faces the so-called ``doublet-triplet'' splitting problem. In the minimal renormalizable DFSZ $\SU(5)$ discussed above the Higgs sector is composed of $\mathbf{5_H}$, $\mathbf{24_H}$, 
$\mathbf{45_H}$ and $\mathbf{5_H^{'}}$. In order to achieve unification in agreement with the low energy values for the gauge couplings and have at least one light Higgs boson, we need to split the $\mathbf{5_H}$ and $\mathbf{45_H}$ 
representations. Unfortunately, in this theory one does not have an explanation to show why some of these fields living in $\mathbf{5_H}$, $\mathbf{45_H}$ and $\mathbf{5_H^{'}}$ are light but we can constrain the Higgs spectrum in the theory 
using all current experimental bounds. We take a phenomenological approach in the sense that we study the experimental implications of the simplest realistic DFSZ $\SU(5)$ model.

\subsection{Axion Mass and Proton Stability}
\label{sec:ProtonDecay}
\begin{figure}[tbp]
\centering
\includegraphics[width=0.6\linewidth]{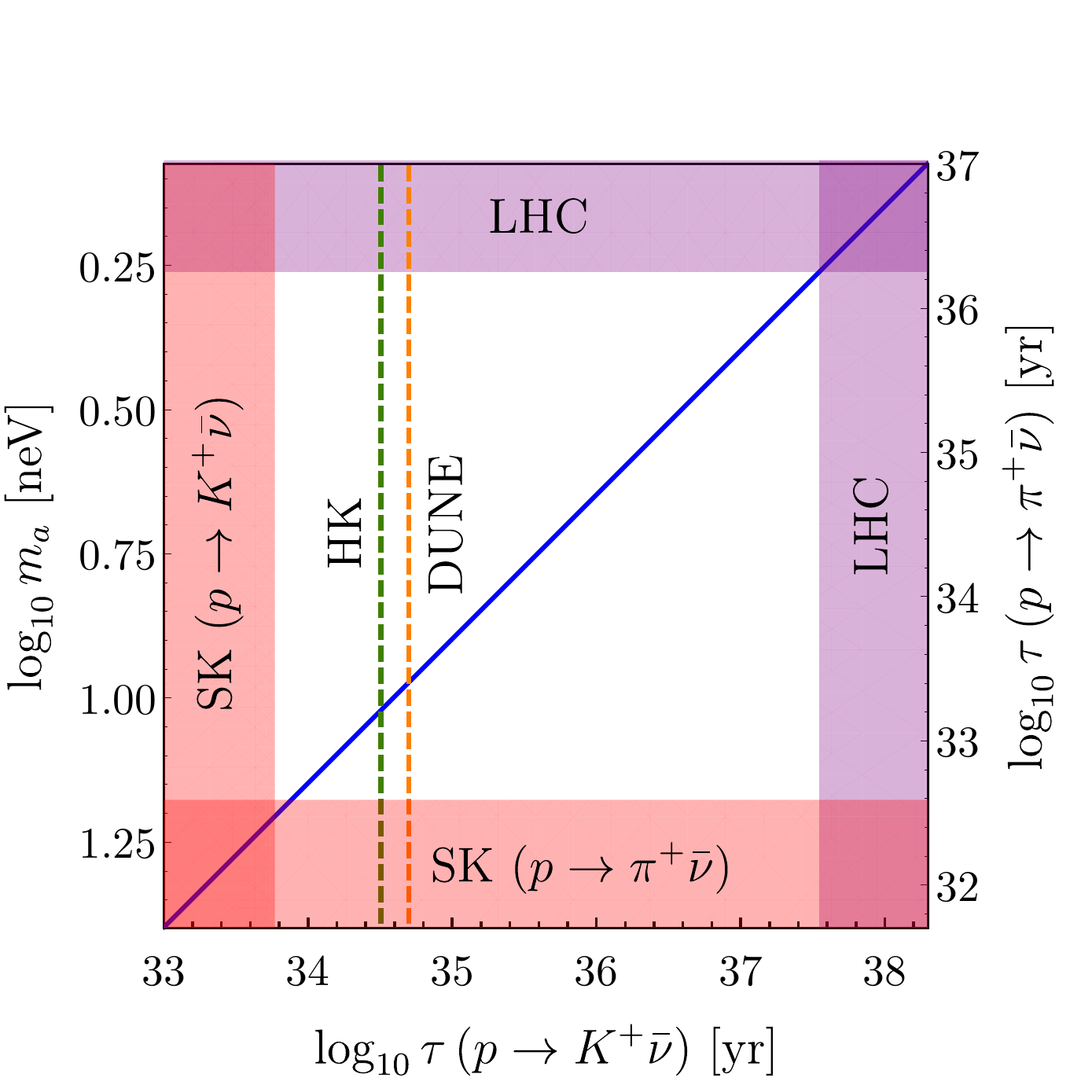}
\caption{Correlation between the proton lifetime prediction for the channels $p \to K^+ \bar{\nu}$ and $p \to \pi^+ \bar{\nu}$ with respect to the axion mass. The regions shaded in red show the parameter space excluded by the proton decay bounds from the Super-Kamiokande collaboration $\tau(p \to K^+ \bar{\nu}) > 5.9 \times 10^{33}$ years~\cite{Abe:2014mwa} and $\tau(p \to \pi^+ \bar{\nu})> 3.9 \times 10^{32}$ years \cite{Abe:2013lua}. The green and orange dashed lines give the projected bounds from the Hyper-Kamiokande collaboration $\tau(p\to K^+ \bar{\nu}) > 3.2 \times 10^{34}$ years~\cite{Abe:2018uyc} and the DUNE collaboration $\tau(p \to K^+ \bar{\nu})> 5 \times 10^{34}$ years~\cite{Acciarri:2015uup}. The region shaded in purple is the parameter space ruled out by collider bounds on $\Phi_1$, i.e. $M_{\Phi_1} > 1$ TeV~\cite{Miralles:2019uzg} assuming $M_{H_{2,3}} = M_\text{GUT}$. The area in white gives the allowed axion mass and proton lifetimes in the context of this theory.}
\label{fig:nudecays}
\end{figure}

%
In this theory the mass matrix for the up-quarks is symmetric and therefore we can predict the decay width for the proton decay channels with antineutrinos as a function of the known mixings at low energy~\cite{FileviezPerez:2004hn}. The decay widths for the $p \to K^+ \bar{\nu}$ and $p \to \pi^+ \bar{\nu}$ channels in the context of this $\SU(5) \otimes \U(1)_\text{PQ}$ theory are given by
\begin{eqnarray}
 \Gamma (p \to K^+ \bar{\nu}) &=&  \frac{\pi m_p}{2} \frac{\alpha_{\rm GUT}^2}{M_{\rm GUT}^4}  \left( 1- \frac{m_{K^+}^2}{m_p^2}\right)^2 C_K, \\
 \frac{ \Gamma (p \to \pi^+ \bar{\nu})}{ \Gamma (p \to K^+ \bar{\nu})} &=&  \frac{C_\pi}{C_K} \left( 1- \frac{m_{K^+}^2}{m_p^2}\right)^{-2}\simeq 20,
\end{eqnarray}
where
\begin{eqnarray}
C_K &=&  A_{RG}^2  \left( \left |V_{\rm CKM}^{11}  
 \matrixel{K^+}{(us)_R d_L}{p} \right |^2 + \left | V_{\rm CKM}^{12}  
 \matrixel{K^+}{(ud)_R s_L}{p} \right|^2 \right) \simeq 0.01 , \\ 
 C_\pi &=&  A_{RG}^2 \left|V_{\rm CKM}^{11} \matrixel{\pi^+}{(du)_R d_L}{p}\right |^2 \simeq 0.1.
\end{eqnarray}
We note that the $C_K$ and $C_\pi$ coefficients are functions of known parameters: the RGE factor $A_{RG}$, which parametrizes the running between the GUT and the $\Lambda_\text{QCD}$ scale~\cite{Nath:2006ut}, and the known values of the CKM matrix. We remark that the calculation of these coefficients in the context of GUTs is only possible if, as in this theory, $M_u = M_u^T$. For the hadronic matrices we use the results from lattice QCD given in Ref.~\cite{Aoki:2017puj}. 
Since the proton decay width and the axion mass both depend on the ratio $(\alpha_{\rm GUT} / M_{\rm GUT}^2)$, we can relate them by
\begin{equation}
m_a \simeq \frac{4.5}{\tau^{1/4}(p\to K^+\bar{\nu})} \text{ eV}.
\end{equation}
Therefore, if any of these two decay channels are discovered in proton decay experiments one can automatically predict the other channel and the axion mass, as shown in Fig.~\ref{fig:nudecays}. In this figure, we present the prediction of the axion mass from the lifetime of any of the proton decay channels into anti-neutrinos. The red shaded area shows the excluded parameter space from Super-Kamiokande (SK) bounds for both $\tau(p \to K^+ \bar{\nu}) > 5.9 \times 10^{33}$ years~\cite{Abe:2014mwa} and $\tau(p \to \pi^+ \bar{\nu})> 3.9 \times 10^{32}$ years \cite{Abe:2013lua}. The projected bounds on the decay channel $p \to K^+ \bar{\nu}$ from the Hyper-Kamiokande collaboration $\tau(p\to K^+ \bar{\nu}) > 3.2 \times 10^{34}$ years~\cite{Abe:2018uyc} and the DUNE collaboration $\tau (p \to K^+ \bar{\nu}) > 5 \times 10^{34}$ years~\cite{Acciarri:2015uup} are shown with a green and orange dashed lines, respectively. The purple shaded areas correspond to the parameter space excluded by collider bounds on the colored doublet $M_{\Phi_1}> 1$ TeV~\cite{Miralles:2019uzg}, where we have assumed the $M_{H_{2,3}}=M_\text{GUT}$ in order to account for the largest possible range. 

The white region in Fig.~\ref{fig:nudecays} shows the available window for the axion mass in this model, which is predicted to be 
\begin{equation}
\label{eq:axionmass}
m_a = (1.87 \, - \, 16.05) \times 10^{-9} \text{ eV}.
\end{equation}
Furthermore, the theory predicts the upper bound on the proton decay lifetime for the channels with antineutrinos, 
\begin{equation}
\tau(p\to K^+ \bar{\nu}) \lesssim   3.5 \times 10^{37} \text{ yr}, \quad  \text{ and }\quad \tau(p \to \pi^+ \bar{\nu}) \lesssim 1.8 \times 10^{36}\text{ yr},
\end{equation}
which expose the theory to be tested in current or future proton decay experiments.
We emphasize that the peculiar feature $M_u = M_u^T$ from this theory allows us to predict the upper bound of the axion mass window.

Unfortunately, the width for the proton decay channel with charged leptons cannot be predicted as a function of known quantities at low energy. The decay width for $p \to \pi^0 e^+$ is given by
\begin{equation}
\Gamma (p \to \pi^0 e^+)=  \frac{\pi m_p}{2} \frac{\alpha_{\rm GUT}^2}{M_{\rm GUT}^4}  A_{RG}^2  V_e^2  \left |\matrixel{\pi^0}{(ud)_R u_L}{p}\right|^2, 
\end{equation}
where $V_e$ is a flavor factor coming from the combination of some unknown fermion mixing matrices. See Appendix~\ref{sec:AppProton} for further details. As we show in that appendix, although the proton lifetime cannot be predicted in this channel, information about the $V_e$ matrix can be inferred from the experimental bounds on proton decay.

\section{Axion phenomenology}
\label{sec:axionpheno}
\begin{figure}[t]
$$\includegraphics[width=0.495\textwidth]{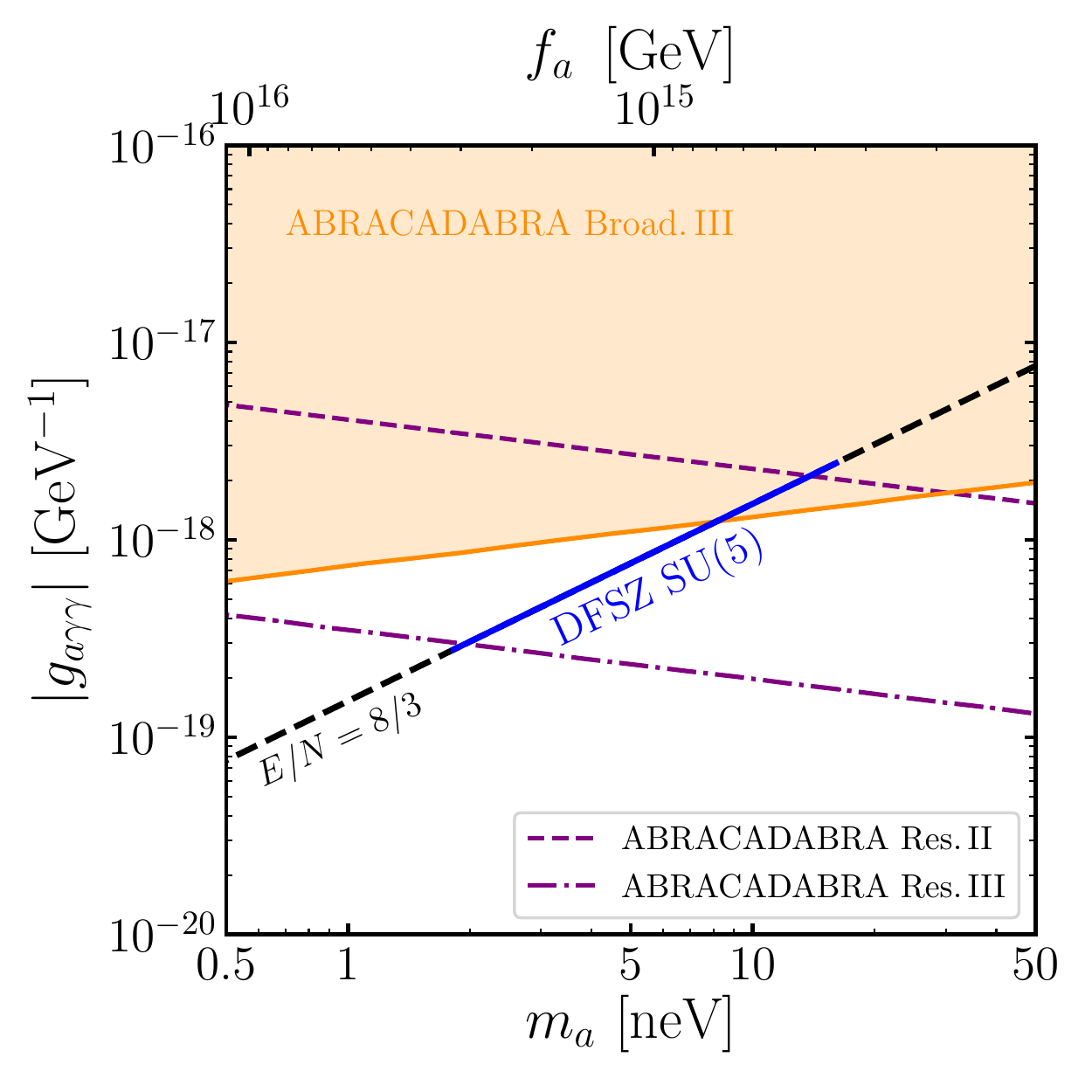} \,\,\,\,
\includegraphics[width=0.495\textwidth]{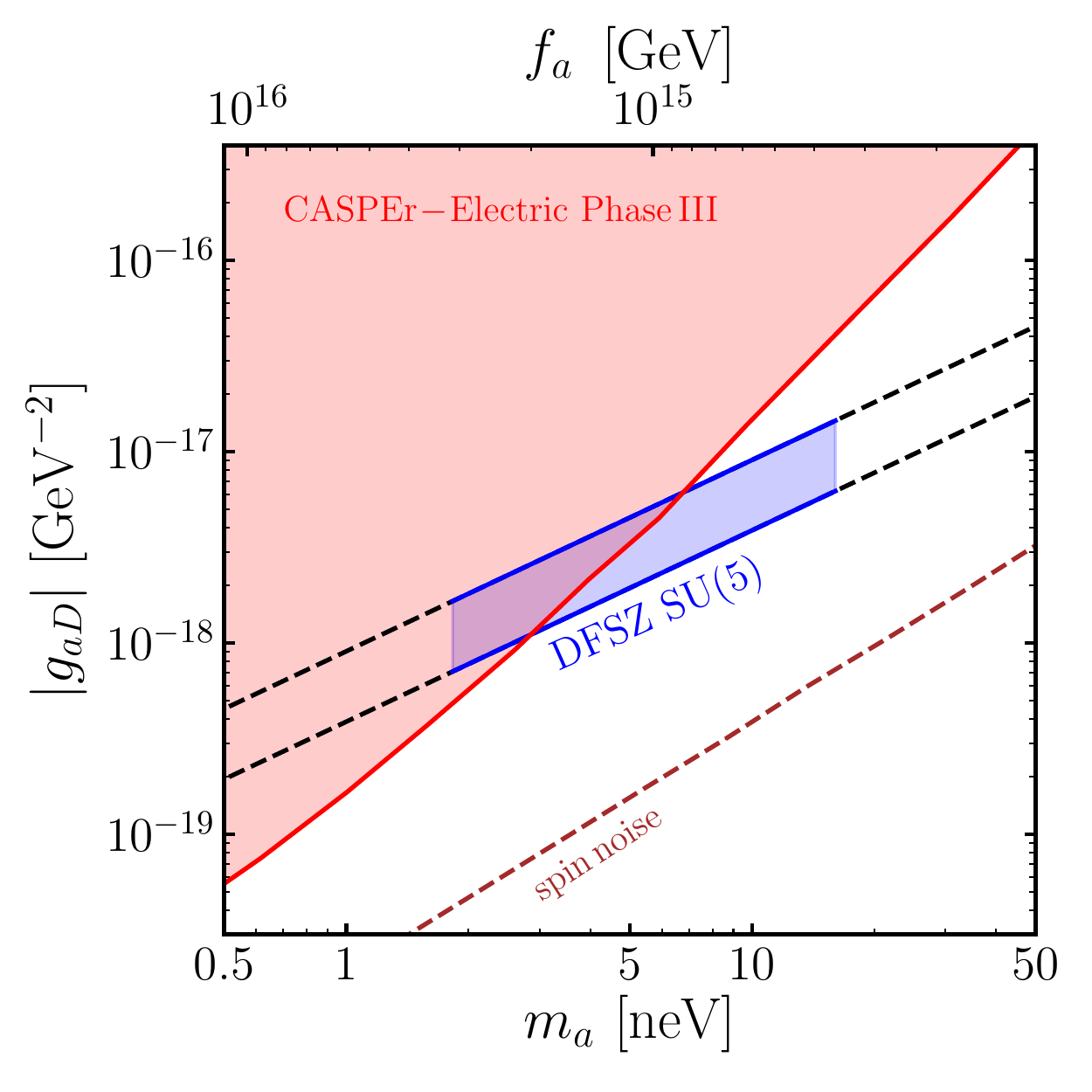} $$
\caption{\textit{Left panel:} The axion coupling to photons as a function of the axion mass. The blue solid line corresponds to the prediction in the theory considered in this paper. The region shaded in orange gives the projected sensitivite for Phase III of the ABRACADABRA~\cite{Kahn:2016aff} experiment using the broadband approach. The purple dashed (dotted-dashed) line corresponds to Phase II (III) of the resonant approach. \textit{Right panel:} The axion coupling to the neutron EDM as a function of the axion mass. The blue band corresponds to the theoretical error on the $g_{aD}$ ocupling~\cite{Pospelov:1999mv}. The region shaded in red gives the projected sensitivity to Phase III of the CASPEr-Electric~\cite{JacksonKimball:2017elr} experiment.}
\label{fig:bounds}
\end{figure}

In this section, we study the axion couplings to SM particles in the predicted mass window. We focus on the axion to photon coupling and the interaction between the axion and the electric dipole moment of the neutron (nEDM) for which there exist experiments that could probe this scenario. In this work, we consider the case in which the PQ symmetry is broken before inflation and in order to achieve the correct dark matter relic abundance we assume an initial misalignment angle of $\theta_i \approx 10^{-2}$ \cite{Preskill:1982cy,Abbott:1982af,Dine:1982ah}.

The interaction between the axion and photons can be obtained by rotating the axion field from the Yukawa terms of the charged fermions
\beq
\mathcal{L} \supset -\frac{g_{a\gamma\gamma}}{4} a F_{\mu\nu} \tilde{F}^{\mu\nu},
\eeq
with the effective coupling $g_{a\gamma\gamma}$ given by
\beq
\label{eq:gagamma}
g_{a\gamma\gamma} = \frac{\alpha_{\tiny{\rm EM}}} {2\pi f_a} \left(\frac{E}{N}  - 1.92(4) \right)  = \frac{\alpha_{\tiny{\rm EM}}} {2\pi f_a} \left(\frac{8}{3}  - 1.92(4) \right),
\eeq
where second term is the contribution from non-perturbative effects from the axion coupling to QCD and has been computed at NLO in Ref.~\cite{diCortona:2015ldu}. 

We present our results in Fig.~\ref{fig:bounds}. The solid blue line corresponds to the $g_{a\gamma\gamma}$ coupling in the predicted mass window and we show the projected sensitivites of the ABRACADABRA experiment~\cite{Kahn:2016aff}. The broadband approach in its Phase III, which corresponds to a configuration with magnetic field of 5 T and a volume of $100 \, {\rm m}^3$, will be sensitive to a portion of this mass window, as shown by the orange region. This sensitivity takes into account only the irreducible source of noise in the experiment. Phase III of the resonant approach, shown by a purple dotted-dashed line, will be able to cover most of the predicted mass window. The latter assumes that the noise in the SQUID is much smaller than the thermal noise. The collaboration has recently published their first results using a prototype detector~\cite{Ouellet:2018beu}. 

The dark matter axion background field induces an oscillating electric dipole moment for the neutron, given by
\beq
d_n = g_{aD} \, a 	\approx 2.4 \times 10^{-16} \frac{a}{f_a} \, e\cdot{\rm cm}.
\eeq 
The CASPEr-Electric \cite{Budker:2013hfa, JacksonKimball:2017elr} experiment aims to measure this oscillating nEDM and Phase III of this experiment will probe the lower portion of the predicted mass window as shown in Fig.~\ref{fig:bounds}. Experimental limits have already been found using this search strategy~\cite{Abel:2017rtm}. However, the advanced stage of CASPEr-Electric we show in Fig.~\ref{fig:bounds} relies on technology that is currently under development. When the projected sensitivities for ABRACADABRA and CASPEr-Electric are combined, these experiments will be able to fully probe the mass window in Eq.~\eqref{eq:axionmass}.

One important difference between this scenario and the one studied in Ref.~\cite{FileviezPerez:2019fku} is that here the axion has a tree-level coupling to electrons. However, all the experimental constraints on this coupling are well above the prediction for the QCD axion with mass around $10^{-9}$ eV. It is important to emphasize that this theory can be fully tested using the predictions for the axion mass and upper bounds on the proton decay lifetimes.
%
\section{Summary}
\label{sec:Summary}
We discussed the implementation of the DFSZ mechanism for the QCD axion mass in the context of grand unified theories.  We have shown that using the idea of Wise, Georgi and Glashow the axion mass can be predicted in a realistic renormalizable grand unified theory. The PQ scale is determined by the GUT scale which allows us to predict the axion mass:
$$m_a \simeq (2 \, - \, 16) \times 10^{-9} \text{ eV}.$$
We have shown that the predictions for the axion couplings can be tested at ABRACADABRA and CASPEr-Electric experiments.  

The fact that the mass matrix for up-quarks is symmetric implies that the proton decay channels with antineutrinos is a function of the known mixings at low energy. In this theory, the upper bounds on the proton decay lifetimes with antineutrinos are given by
$$\tau(p\to K^+ \bar{\nu}) \lesssim   4 \times 10^{37} \text{ yr},  \quad \text{ and }\quad \tau(p \to \pi^+ \bar{\nu}) \lesssim 2 \times 10^{36}\text{ yr}.$$
This theory is unique due to the fact that it can be fully probed by proton decay experiments such as DUNE and axion experiments such as ABRACADABRA.\\

{\textit{Acknowledgments}}:
We thank Goran Senjanovi\'c for a wonderful discussion about the strong CP problem.
P.F.P. and C.M. thank Mark B. Wise for many discussions about models for axions and the theory group at Caltech for hospitality. The work of P.F.P. has been supported by the U.S. Department of Energy, Office of Science, Office of High Energy Physics, under Award Number DE-SC0020443. The work of C.M. has been supported in part by Grants No. FPA2014-53631-C2-1-P, No. FPA2017-84445-P, 
and No. SEV-2014- 0398 (AEI/ERDF, EU), and by a La Caixa-Severo Ochoa scholarship. 

\appendix
\newpage

\section{Proton Decay: $p \to \pi^0 e^+$}
\label{sec:AppProton}
\begin{figure}[h]
\centering
\includegraphics[height=7cm]{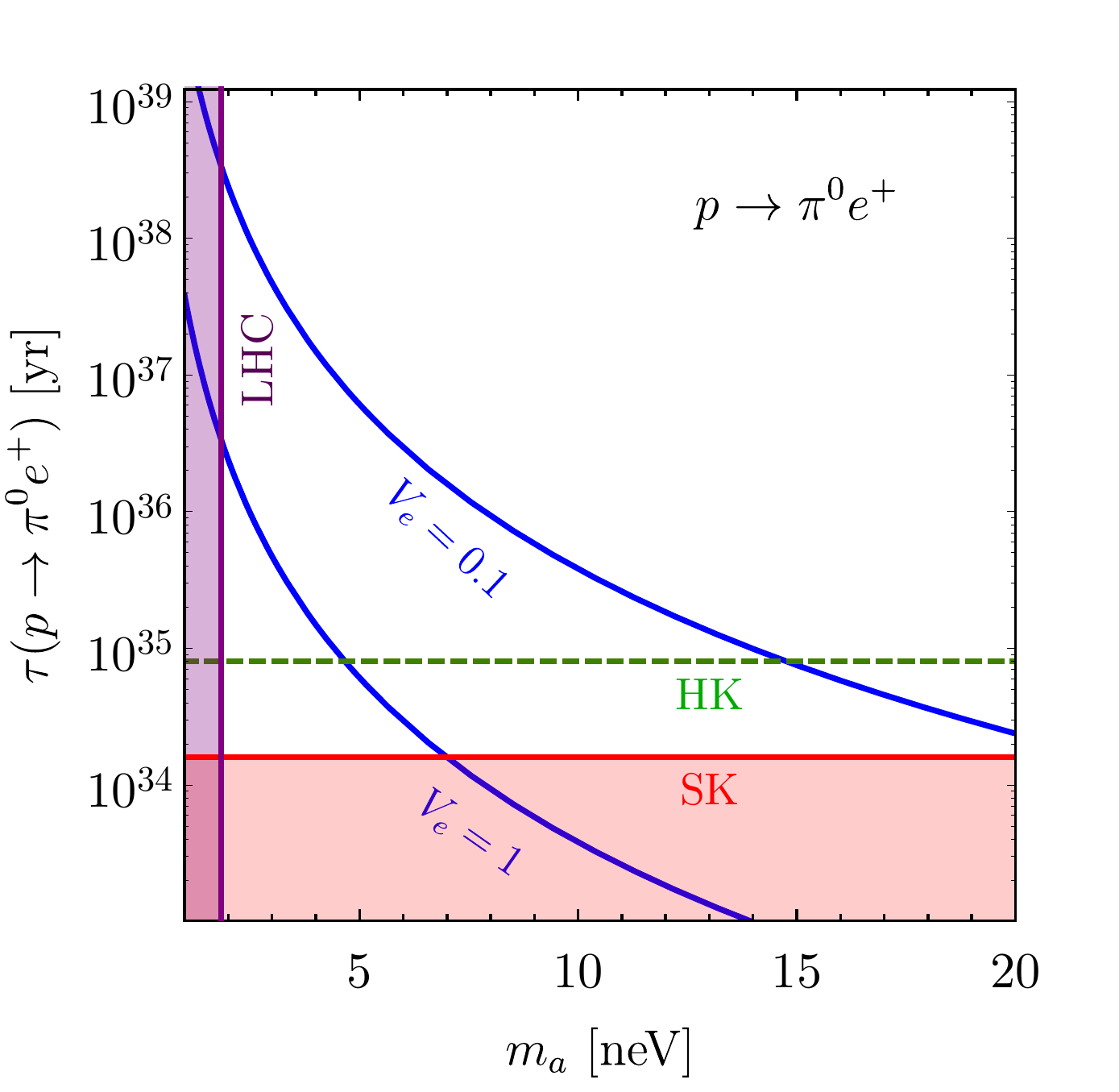}
\caption{Predictions for the proton lifetime from the decay channel $p \to \pi^0 e^+$. The red area shows the Super-Kamiokande constraints,  $\tau ( p\to \pi^0 e^+) > 1.6 \times 10^{34}$ years~\cite{Miura:2016krn}, 
while the green dashed line shows the Hyper-Kamiokande collaboration projected bound, $\tau(p \to \pi^0 e^+) > 8 \times 10^{34}$ years~\cite{Yokoyama:2017mnt}. The region shaded in purple is ruled out by the collider constraints on the $\Phi_1$ field, $M_{\Phi_1} > 1 $ TeV~\cite{Miralles:2019uzg}.}
\label{fig:nudecayspi0}
\end{figure}

%
The decay rate for the proton decay channel $p \to \pi^0 e^+$ is given by
\begin{eqnarray}
\Gamma (p \to \pi^0 e^+)&=&   \frac{\pi m_p}{2} \frac{\alpha_{\rm GUT}^2}{M_{\rm GUT}^4}  A_{RG}^2  V_e^2  \left |\matrixel{\pi^0}{(ud)_R u_L}{p}\right|^2,
\end{eqnarray}
where
\begin{equation}
V_e= \sqrt{ \left(  \left| (V_2^{11} +V_\text{CKM}^{11}(V_2 V_\text{CKM})^{11})\right |^2 + \left |V_3^{11} \right |^2 \right)}.
\end{equation}
In Fig.~\ref{fig:nudecayspi0} we show the predictions for $p \to \pi^0 e^+$,  together with the Super-Kamiokande constraints (red area), $\tau ( p\to \pi^0 e^+) > 1.6 \times 10^{34}$ years~\cite{Miura:2016krn}, 
and the Hyper-Kamiokande projected bound (green line), $\tau(p \to \pi^0 e^+) > 8 \times 10^{34}$ years~\cite{Yokoyama:2017mnt}. To illustrate the effect of the unkown matrix $V_e$, we show the numerical predictions for two possible 
values of this matrix $V_e = 0.1$ and $V_e=1$. Notice that the matrix $V_e$ can be constrained using the proton decay experimental bounds for the decay into charged leptons, but in general  the lifetime for this channel cannot be predicted.

\newpage
\section{Neutrino masses}
\label{sec:AppNu}
The theory we have discussed so far does not have a mechanism to generate neutrino masses. However, this can be addressed by adding  three neutrinos singlets, $\nu^C_i$, with a Majorana mass term and the interaction $Y_\nu \, \bar{5} \, 5_H \nu^C$, then we can implement the type I seesaw mechanism~\cite{Minkowski:1977sc,Mohapatra:1979ia,GellMann:1980vs,Yanagida:1979as}. This will fix the PQ charges to $\beta\!=\!-2\alpha$, or $\beta\!=\!\alpha/2$ if we use $\mathbf{5_H'}$ instead of $\mathbf{5_H}$. The baryon asymmetry of the Universe can then be explained through thermal leptogenesis~\cite{Fukugita:1986hr} or through out-of-equilibrium decays of the heavy colored Higgs~\cite{Fukugita:2002hu}.


An alternative is to introduce a $\mathbf{10_H}$~\cite{Perez:2016qbo} and implement the Zee mechanism~\cite{Zee:1980ai}, in which neutrinos acquire mass at the radiative level. In this scenario, the relevant Lagrangian for neutrino masses is given by
\begin{equation}
{\cal L} \supset \lambda \ \bar{5} \ \bar{5}  \ 10_H + 10 \ \bar{5}  \left( Y_1 \ 5_H^* + Y_2 \ 45_H^* \right) + \mu \ 5_H 45_H  10_H^* + \text{h.c.}.
\end{equation}
We note that the Higgs in the $\mathbf{45_H}$ plays a twofold role: it corrects the mass relation between charged leptons and down-type quarks and contributes to the generation of neutrino masses at the quantum level. We also note that the implementation of this mechanism in the renormalizable $\SU(5)$ we proposed would fix all relative Peccei-Quinn charges among the field content, i.e. 
\begin{eqnarray}
&& \bar{5} \to e^{i \alpha} \ \bar{5}, \hspace{2.5cm}  10 \to  e^{ - 2i \alpha} \ 10, \hspace{2.4cm}   5_H' \to e^{ 4i\alpha} \  5_H', \nonumber \\
&& 24_H \to e^{ 5i \alpha/2 } \ 24_H,  \hspace{1.15cm}   5_H / 45_H \to e^{ -i \alpha }\  5_H/45_H, \hspace{0.62cm}  10_H  \to e^{-2i\alpha} \ 10_H. \nonumber
\end{eqnarray}
These are basically the simplest possibilities to generate neutrino masses when implementing only the DFSZ mechanism for the axion mass. See our recent study in Ref.~\cite{FileviezPerez:2019fku} for the discussion of other possibilities.

\bibliographystyle{JHEP}
\bibliography{axions}{}

\providecommand{\href}[2]{#2}\begingroup\raggedright\begin{thebibliography}{10}

\bibitem{Peccei:1977hh}
R.~D. Peccei and H.~R. Quinn, \emph{{CP Conservation in the Presence of
  Instantons}}, \href{https://doi.org/10.1103/PhysRevLett.38.1440}{\emph{Phys.
  Rev. Lett.} {\bfseries 38} (1977) 1440--1443}.

\bibitem{Wilczek:1977pj}
F.~Wilczek, \emph{{Problem of Strong $P$ and $T$ Invariance in the Presence of
  Instantons}}, \href{https://doi.org/10.1103/PhysRevLett.40.279}{\emph{Phys.
  Rev. Lett.} {\bfseries 40} (1978) 279--282}.

\bibitem{Weinberg:1977ma}
S.~Weinberg, \emph{{A New Light Boson?}},
  \href{https://doi.org/10.1103/PhysRevLett.40.223}{\emph{Phys. Rev. Lett.}
  {\bfseries 40} (1978) 223--226}.

\bibitem{Preskill:1982cy}
J.~Preskill, M.~B. Wise and F.~Wilczek, \emph{{Cosmology of the Invisible
  Axion}}, \href{https://doi.org/10.1016/0370-2693(83)90637-8}{\emph{Phys.
  Lett.} {\bfseries B120} (1983) 127--132}.

\bibitem{Abbott:1982af}
L.~F. Abbott and P.~Sikivie, \emph{{A Cosmological Bound on the Invisible
  Axion}}, \href{https://doi.org/10.1016/0370-2693(83)90638-X}{\emph{Phys.
  Lett.} {\bfseries B120} (1983) 133--136}.

\bibitem{Dine:1982ah}
M.~Dine and W.~Fischler, \emph{{The Not So Harmless Axion}},
  \href{https://doi.org/10.1016/0370-2693(83)90639-1}{\emph{Phys. Lett.}
  {\bfseries B120} (1983) 137--141}.

\bibitem{Raffelt:1990yz}
G.~G. Raffelt, \emph{{Astrophysical methods to constrain axions and other novel
  particle phenomena}},
  \href{https://doi.org/10.1016/0370-1573(90)90054-6}{\emph{Phys. Rept.}
  {\bfseries 198} (1990) 1--113}.

\bibitem{Dine:2000cj}
M.~Dine, \emph{{TASI lectures on the strong CP problem}},  in \emph{{Flavor
  physics for the millennium. Proceedings, Theoretical Advanced Study Institute
  in elementary particle physics, TASI 2000, Boulder, USA, June 4-30, 2000}},
  pp.~349--369, 2000, \href{https://arxiv.org/abs/hep-ph/0011376}{{\ttfamily
  hep-ph/0011376}}.

\bibitem{Sikivie:2006ni}
P.~Sikivie, \emph{{Axion Cosmology}},
  \href{https://doi.org/10.1007/978-3-540-73518-2_2}{\emph{Lect. Notes Phys.}
  {\bfseries 741} (2008) 19--50},
  [\href{https://arxiv.org/abs/astro-ph/0610440}{{\ttfamily
  astro-ph/0610440}}].

\bibitem{Kim:2008hd}
J.~E. Kim and G.~Carosi, \emph{{Axions and the Strong CP Problem}},
  \href{https://doi.org/10.1103/RevModPhys.82.557}{\emph{Rev. Mod. Phys.}
  {\bfseries 82} (2010) 557--602},
  [\href{https://arxiv.org/abs/0807.3125}{{\ttfamily 0807.3125}}].

\bibitem{Jaeckel:2010ni}
J.~Jaeckel and A.~Ringwald, \emph{{The Low-Energy Frontier of Particle
  Physics}},
  \href{https://doi.org/10.1146/annurev.nucl.012809.104433}{\emph{Ann. Rev.
  Nucl. Part. Sci.} {\bfseries 60} (2010) 405--437},
  [\href{https://arxiv.org/abs/1002.0329}{{\ttfamily 1002.0329}}].

\bibitem{Marsh:2015xka}
D.~J.~E. Marsh, \emph{{Axion Cosmology}},
  \href{https://doi.org/10.1016/j.physrep.2016.06.005}{\emph{Phys. Rept.}
  {\bfseries 643} (2016) 1--79},
  [\href{https://arxiv.org/abs/1510.07633}{{\ttfamily 1510.07633}}].

\bibitem{Graham:2015ouw}
P.~W. Graham, I.~G. Irastorza, S.~K. Lamoreaux, A.~Lindner and K.~A. van
  Bibber, \emph{{Experimental Searches for the Axion and Axion-Like
  Particles}},
  \href{https://doi.org/10.1146/annurev-nucl-102014-022120}{\emph{Ann. Rev.
  Nucl. Part. Sci.} {\bfseries 65} (2015) 485--514},
  [\href{https://arxiv.org/abs/1602.00039}{{\ttfamily 1602.00039}}].

\bibitem{Irastorza:2018dyq}
I.~G. Irastorza and J.~Redondo, \emph{{New experimental approaches in the
  search for axion-like particles}},
  \href{https://doi.org/10.1016/j.ppnp.2018.05.003}{\emph{Prog. Part. Nucl.
  Phys.} {\bfseries 102} (2018) 89--159},
  [\href{https://arxiv.org/abs/1801.08127}{{\ttfamily 1801.08127}}].

\bibitem{Zhitnitsky:1980tq}
A.~R. Zhitnitsky, \emph{{On Possible Suppression of the Axion Hadron
  Interactions. (In Russian)}}, {\emph{Sov. J. Nucl. Phys.} {\bfseries 31}
  (1980) 260}.

\bibitem{Dine:1981rt}
M.~Dine, W.~Fischler and M.~Srednicki, \emph{{A Simple Solution to the Strong
  CP Problem with a Harmless Axion}},
  \href{https://doi.org/10.1016/0370-2693(81)90590-6}{\emph{Phys. Lett.}
  {\bfseries 104B} (1981) 199--202}.

\bibitem{Kim:1979if}
J.~E. Kim, \emph{{Weak Interaction Singlet and Strong CP Invariance}},
  \href{https://doi.org/10.1103/PhysRevLett.43.103}{\emph{Phys. Rev. Lett.}
  {\bfseries 43} (1979) 103}.

\bibitem{Shifman:1979if}
M.~A. Shifman, A.~I. Vainshtein and V.~I. Zakharov, \emph{{Can Confinement
  Ensure Natural CP Invariance of Strong Interactions?}},
  \href{https://doi.org/10.1016/0550-3213(80)90209-6}{\emph{Nucl. Phys.}
  {\bfseries B166} (1980) 493--506}.

\bibitem{FileviezPerez:2019fku}
P.~Fileviez~Pérez, C.~Murgui and A.~D. Plascencia, \emph{{The QCD Axion and
  Unification}}, \href{https://doi.org/10.1007/JHEP11(2019)093}{\emph{JHEP}
  {\bfseries 11} (2019) 093},
  [\href{https://arxiv.org/abs/1908.01772}{{\ttfamily 1908.01772}}].

\bibitem{Co:2016vsi}
R.~T. Co, F.~D'Eramo and L.~J. Hall, \emph{{Supersymmetric axion grand unified
  theories and their predictions}},
  \href{https://doi.org/10.1103/PhysRevD.94.075001}{\emph{Phys. Rev.}
  {\bfseries D94} (2016) 075001},
  [\href{https://arxiv.org/abs/1603.04439}{{\ttfamily 1603.04439}}].

\bibitem{Boucenna:2017fna}
S.~M. Boucenna and Q.~Shafi, \emph{{Axion inflation, proton decay, and
  leptogenesis in $SU(5)\times U(1)_{PQ}$}},
  \href{https://doi.org/10.1103/PhysRevD.97.075012}{\emph{Phys. Rev.}
  {\bfseries D97} (2018) 075012},
  [\href{https://arxiv.org/abs/1712.06526}{{\ttfamily 1712.06526}}].

\bibitem{DiLuzio:2018gqe}
L.~Di~Luzio, A.~Ringwald and C.~Tamarit, \emph{{Axion mass prediction from
  minimal grand unification}},
  \href{https://doi.org/10.1103/PhysRevD.98.095011}{\emph{Phys. Rev.}
  {\bfseries D98} (2018) 095011},
  [\href{https://arxiv.org/abs/1807.09769}{{\ttfamily 1807.09769}}].

\bibitem{Ernst:2018bib}
A.~Ernst, A.~Ringwald and C.~Tamarit, \emph{{Axion Predictions in $SO(10)\times
  U(1)_{\rm PQ}$ Models}},
  \href{https://doi.org/10.1007/JHEP02(2018)103}{\emph{JHEP} {\bfseries 02}
  (2018) 103}, [\href{https://arxiv.org/abs/1801.04906}{{\ttfamily
  1801.04906}}].

\bibitem{Wise:1981ry}
M.~B. Wise, H.~Georgi and S.~L. Glashow, \emph{{SU(5) and the Invisible
  Axion}}, \href{https://doi.org/10.1103/PhysRevLett.47.402}{\emph{Phys. Rev.
  Lett.} {\bfseries 47} (1981) 402}.

\bibitem{Kahn:2016aff}
Y.~Kahn, B.~R. Safdi and J.~Thaler, \emph{{Broadband and Resonant Approaches to
  Axion Dark Matter Detection}},
  \href{https://doi.org/10.1103/PhysRevLett.117.141801}{\emph{Phys. Rev. Lett.}
  {\bfseries 117} (2016) 141801},
  [\href{https://arxiv.org/abs/1602.01086}{{\ttfamily 1602.01086}}].

\bibitem{Budker:2013hfa}
D.~Budker, P.~W. Graham, M.~Ledbetter, S.~Rajendran and A.~Sushkov,
  \emph{{Proposal for a Cosmic Axion Spin Precession Experiment (CASPEr)}},
  \href{https://doi.org/10.1103/PhysRevX.4.021030}{\emph{Phys. Rev.} {\bfseries
  X4} (2014) 021030}, [\href{https://arxiv.org/abs/1306.6089}{{\ttfamily
  1306.6089}}].

\bibitem{FileviezPerez:2004hn}
P.~Fileviez~Perez, \emph{{Fermion mixings versus d = 6 proton decay}},
  \href{https://doi.org/10.1016/j.physletb.2004.06.061}{\emph{Phys. Lett.}
  {\bfseries B595} (2004) 476--483},
  [\href{https://arxiv.org/abs/hep-ph/0403286}{{\ttfamily hep-ph/0403286}}].

\bibitem{Gorghetto:2018ocs}
M.~Gorghetto and G.~Villadoro, \emph{{Topological Susceptibility and QCD Axion
  Mass: QED and NNLO corrections}},
  \href{https://doi.org/10.1007/JHEP03(2019)033}{\emph{JHEP} {\bfseries 03}
  (2019) 033}, [\href{https://arxiv.org/abs/1812.01008}{{\ttfamily
  1812.01008}}].

\bibitem{Tanabashi:2018oca}
{\scshape Particle Data Group} collaboration, M.~Tanabashi et~al.,
  \emph{{Review of Particle Physics}},
  \href{https://doi.org/10.1103/PhysRevD.98.030001}{\emph{Phys. Rev.}
  {\bfseries D98} (2018) 030001}.

\bibitem{Giveon:1991zm}
A.~Giveon, L.~J. Hall and U.~Sarid, \emph{{SU(5) unification revisited}},
  \href{https://doi.org/10.1016/0370-2693(91)91289-8}{\emph{Phys. Lett.}
  {\bfseries B271} (1991) 138--144}.

\bibitem{Miralles:2019uzg}
V.~Miralles and A.~Pich, \emph{{LHC bounds on coloured scalars}},
  \href{https://arxiv.org/abs/1910.07947}{{\ttfamily 1910.07947}}.

\bibitem{Abe:2014mwa}
{\scshape Super-Kamiokande} collaboration, K.~Abe et~al., \emph{{Search for
  proton decay via $p\to \bar{\nu} K^+$ using 260 kiloton·year data of
  Super-Kamiokande}},
  \href{https://doi.org/10.1103/PhysRevD.90.072005}{\emph{Phys. Rev.}
  {\bfseries D90} (2014) 072005},
  [\href{https://arxiv.org/abs/1408.1195}{{\ttfamily 1408.1195}}].

\bibitem{Abe:2018uyc}
{\scshape Hyper-Kamiokande} collaboration, K.~Abe et~al.,
  \emph{{Hyper-Kamiokande Design Report}},
  \href{https://arxiv.org/abs/1805.04163}{{\ttfamily 1805.04163}}.

\bibitem{Acciarri:2015uup}
{\scshape DUNE} collaboration, R.~Acciarri et~al., \emph{{Long-Baseline
  Neutrino Facility (LBNF) and Deep Underground Neutrino Experiment (DUNE)}},
  \href{https://arxiv.org/abs/1512.06148}{{\ttfamily 1512.06148}}.

\bibitem{Nath:2006ut}
P.~Nath and P.~Fileviez~Perez, \emph{{Proton stability in grand unified
  theories, in strings and in branes}},
  \href{https://doi.org/10.1016/j.physrep.2007.02.010}{\emph{Phys. Rept.}
  {\bfseries 441} (2007) 191--317},
  [\href{https://arxiv.org/abs/hep-ph/0601023}{{\ttfamily hep-ph/0601023}}].

\bibitem{Aoki:2017puj}
Y.~Aoki, T.~Izubuchi, E.~Shintani and A.~Soni, \emph{{Improved lattice
  computation of proton decay matrix elements}},
  \href{https://doi.org/10.1103/PhysRevD.96.014506}{\emph{Phys. Rev.}
  {\bfseries D96} (2017) 014506},
  [\href{https://arxiv.org/abs/1705.01338}{{\ttfamily 1705.01338}}].

\bibitem{Abe:2013lua}
{\scshape Super-Kamiokande} collaboration, K.~Abe et~al., \emph{{Search for
  Nucleon Decay via $n \to \bar{\nu} \pi^{0}$ and $p \to \bar{\nu} \pi^{+}$ in
  Super-Kamiokande}},
  \href{https://doi.org/10.1103/PhysRevLett.113.121802}{\emph{Phys. Rev. Lett.}
  {\bfseries 113} (2014) 121802},
  [\href{https://arxiv.org/abs/1305.4391}{{\ttfamily 1305.4391}}].

\bibitem{Pospelov:1999mv}
M.~Pospelov and A.~Ritz, \emph{{Theta vacua, QCD sum rules, and the neutron
  electric dipole moment}},
  \href{https://doi.org/10.1016/S0550-3213(99)00817-2}{\emph{Nucl. Phys.}
  {\bfseries B573} (2000) 177--200},
  [\href{https://arxiv.org/abs/hep-ph/9908508}{{\ttfamily hep-ph/9908508}}].

\bibitem{JacksonKimball:2017elr}
D.~F. Jackson~Kimball et~al., \emph{{Overview of the Cosmic Axion Spin
  Precession Experiment (CASPEr)}},
  \href{https://arxiv.org/abs/1711.08999}{{\ttfamily 1711.08999}}.

\bibitem{diCortona:2015ldu}
G.~Grilli~di Cortona, E.~Hardy, J.~Pardo~Vega and G.~Villadoro, \emph{{The QCD
  axion, precisely}},
  \href{https://doi.org/10.1007/JHEP01(2016)034}{\emph{JHEP} {\bfseries 01}
  (2016) 034}, [\href{https://arxiv.org/abs/1511.02867}{{\ttfamily
  1511.02867}}].

\bibitem{Ouellet:2018beu}
J.~L. Ouellet et~al., \emph{{First Results from ABRACADABRA-10 cm: A Search for
  Sub-$\mu$eV Axion Dark Matter}},
  \href{https://doi.org/10.1103/PhysRevLett.122.121802}{\emph{Phys. Rev. Lett.}
  {\bfseries 122} (2019) 121802},
  [\href{https://arxiv.org/abs/1810.12257}{{\ttfamily 1810.12257}}].

\bibitem{Abel:2017rtm}
C.~Abel et~al., \emph{{Search for Axionlike Dark Matter through Nuclear Spin
  Precession in Electric and Magnetic Fields}},
  \href{https://doi.org/10.1103/PhysRevX.7.041034}{\emph{Phys. Rev.} {\bfseries
  X7} (2017) 041034}, [\href{https://arxiv.org/abs/1708.06367}{{\ttfamily
  1708.06367}}].

\bibitem{Miura:2016krn}
{\scshape Super-Kamiokande} collaboration, K.~Abe et~al., \emph{{Search for
  proton decay via $p \to e^+\pi^0$ and $p \to \mu^+\pi^0$ in 0.31
  megaton·years exposure of the Super-Kamiokande water Cherenkov detector}},
  \href{https://doi.org/10.1103/PhysRevD.95.012004}{\emph{Phys. Rev.}
  {\bfseries D95} (2017) 012004},
  [\href{https://arxiv.org/abs/1610.03597}{{\ttfamily 1610.03597}}].

\bibitem{Yokoyama:2017mnt}
{\scshape Hyper-Kamiokande Proto} collaboration, M.~Yokoyama, \emph{{The
  Hyper-Kamiokande Experiment}},  in \emph{{Proceedings, Prospects in Neutrino
  Physics (NuPhys2016): London, UK, December 12-14, 2016}}, 2017,
  \href{https://arxiv.org/abs/1705.00306}{{\ttfamily 1705.00306}}.

\bibitem{Minkowski:1977sc}
P.~Minkowski, \emph{{$\mu \to e\gamma$ at a Rate of One Out of $10^{9}$ Muon
  Decays?}}, \href{https://doi.org/10.1016/0370-2693(77)90435-X}{\emph{Phys.
  Lett.} {\bfseries 67B} (1977) 421--428}.

\bibitem{Mohapatra:1979ia}
R.~N. Mohapatra and G.~Senjanovic, \emph{{Neutrino Mass and Spontaneous Parity
  Nonconservation}},
  \href{https://doi.org/10.1103/PhysRevLett.44.912}{\emph{Phys. Rev. Lett.}
  {\bfseries 44} (1980) 912}.

\bibitem{GellMann:1980vs}
M.~Gell-Mann, P.~Ramond and R.~Slansky, \emph{{Complex Spinors and Unified
  Theories}}, {\emph{Conf. Proc.} {\bfseries C790927} (1979) 315--321},
  [\href{https://arxiv.org/abs/1306.4669}{{\ttfamily 1306.4669}}].

\bibitem{Yanagida:1979as}
T.~Yanagida, \emph{{Horizontal gauge symmetry and masses of neutrinos}},
  {\emph{Conf. Proc.} {\bfseries C7902131} (1979) 95--99}.

\bibitem{Fukugita:1986hr}
M.~Fukugita and T.~Yanagida, \emph{{Baryogenesis Without Grand Unification}},
  \href{https://doi.org/10.1016/0370-2693(86)91126-3}{\emph{Phys. Lett.}
  {\bfseries B174} (1986) 45--47}.

\bibitem{Fukugita:2002hu}
M.~Fukugita and T.~Yanagida, \emph{{Resurrection of grand unified theory
  baryogenesis}},
  \href{https://doi.org/10.1103/PhysRevLett.89.131602}{\emph{Phys. Rev. Lett.}
  {\bfseries 89} (2002) 131602},
  [\href{https://arxiv.org/abs/hep-ph/0203194}{{\ttfamily hep-ph/0203194}}].

\bibitem{Perez:2016qbo}
P.~Fileviez~Perez and C.~Murgui, \emph{{Renormalizable SU(5) Unification}},
  \href{https://doi.org/10.1103/PhysRevD.94.075014}{\emph{Phys. Rev.}
  {\bfseries D94} (2016) 075014},
  [\href{https://arxiv.org/abs/1604.03377}{{\ttfamily 1604.03377}}].

\bibitem{Zee:1980ai}
A.~Zee, \emph{{A Theory of Lepton Number Violation, Neutrino Majorana Mass, and
  Oscillation}}, \href{https://doi.org/10.1016/0370-2693(80)90349-4,
  10.1016/0370-2693(80)90193-8}{\emph{Phys. Lett.} {\bfseries 93B} (1980) 389}.

\end{thebibliography}\endgroup
\end{document}